# 3D stochastic interferometer detects picometer deformations and minute dielectric fluctuations of its optical volume

**Authors:** Guillaume Graciani[1,2], Marcel Filoche[3], François Amblard[2,4*]

**Affiliations:**

[1] Institute for Molecular Biology and Genetics, Seoul National University, Seoul 08826, South Korea

[2] Institute for Basic Science, Center for Soft and Living Matter, Ulsan, South Korea.

[3] Laboratoire de Physique de la Matière Condensée, Ecole Polytechnique, CNRS, IP Paris, Palaiseau, France.

[4] Department of Physics, Ulsan National Institute of Science and Technology, Ulsan, South Korea.

*Correspondence to: famblard@protonmail.com.

**Abstract**

Optical interferometry has proven extremely powerful to investigate some of the most fundamental laws of physics, using light with very precisely controlled geometry. In the past few decades, various speckle metrology methods have emerged to harness the interferometric properties of strongly disordered light instead, using time-domain analysis of speckle patterns at a maximum rate limited by the frequency of image acquisition.

The present work is based on using a centimeter-sized quartz-powder cavity with arbitrary shape that is engineered with very high Lambertian reflectivity. When filled with a coherent monochromatic photon gas, a statistically isotropic and homogeneous 3D speckle interference pattern is obtained. A single-mode fiber is then used in combination with photon number autocorrelation analysis to detect minute changes of the speckle decorrelation spectrum, caused either by cavity deformations or fluctuations of the dielectric tensor field inside. Due to the statistical homogeneity of the field, the interferometric response to such causes is non-local, since it depends neither on where the cause sits nor where the response is measured. The decorrelation spectrum is acquired over 8 to 10 frequency decades below 100 MHz, with a sensitivity that is only limited by the intrinsic photon statistics and the extrinsic instrumental noises.

With typically 1700 reflections and an average photon transit path length of 62m, our 3D stochastic interferometer yields a typical finesse of 10500, and cavity deformations are detected in an ergodic fashion over a six-decade dynamic range with a power noise floor of $4 10^{-3}$ pm$^2$ which corresponds to 2.7 pm at 1 kHz.

The cavity also operates as a speckle fluctuation amplifier that reveals decorrelation spectra due to picometric thermal motions of colloids in both single and multiple scattering regimes with a typical 100-fold sensitivity gain compared to conventional light scattering techniques.

**Introduction**

Light interferometry is most familiar to us as a unique tool for measuring variations of optical lengths or wavelengths, *i.e.* 1D optical features. From Fizeau's experiment in 1851 on the speed of light *(1)* or Michelson and Morley's measurement of the relative motion of the Earth and the Luminiferous Ether in 1887 *(2)*, to the most recent detection of gravitational waves down to typical strains of $10^{-21}$*(3)* provided by the LIGO and VIRGO observatories, optical interferometers have consistently redefined what constitutes the limits of metrology and brought major contributions to our understanding of physical laws.

Unlike these techniques which require an extremely precise geometry of the optical field, speckle metrology methods have been devised in the past decades to harness the interferometric response of strongly disordered but coherent light. While it has proven to provide exceptionally sensitive measurements, in particular for wavelength determination and spectral analysis *(4,5)*, it has also been used recently for small angles and displacements metrology *(6)* or refractive index sensing *(7)*, and recent quantitative comparisons indicate that it can reach sensitivity levels that come close to those of interferometers using well-ordered light *(8)*.

However and to the best of our knowledge, all speckle metrology methods are based on the time-domain analysis of speckle patterns, *i.e.* of the images of the speckle field, either through the analysis of spatial correlations of the interference pattern *(8)* or even the analysis of spatial phase singularities in the field *(9)*. Consequently, time-frequencies cannot be probed beyond the image acquisition rate.

In the present work, we propose instead to measure the intensity fluctuation of a single speckle grain to calculate the frequency spectrum of perturbations. By homogeneously filling an ultra-high reflectivity Lambertian cavity of arbitrary shape with a coherent and isotropic photon gas, we produce a fully developed 3D speckle field that is ergodically sensitive to minute fluctuations of its optical volume. Here, we present the design of this 3D stochastic interferometer, together with theoretical predictions of sensitivity confirmed by experimental results. We find that the frequency spectrum is sensitive to volume deformations across 8 to 10 frequency decades below 100 MHz, with a sensitivity better than $10^{-5}\lambda$ at 1 kHz. We also show that the interferometer can measure picometric motions of scattering objects placed inside the cavity. We show an increase in sensitivity of about 100-folds in comparison to conventional and multiple light scattering techniques.

**Results**

**Description of a 3D random light field**

Practically, a single-frequency laser light with central wavelength $\lambda_0 = 660nm$ and coherence length $\Lambda_{coh} \approx 95m$ is fiber-coupled into a closed diffuse reflective cavity, with walls made of compressed quartz powder *(10)* that provide a uniform Lambertian reflectivity with an albedo $\mathcal{A}$ close to unity. A cylindrical shape was used with centimetric size (Fig.1A). Thanks to a very small reflection loss coefficient $\epsilon = 1 - \mathcal{A} \approx 6 \pm 0.5.10^{-4}$, each photon is subjected to an average number of reflections given by $G_w = 1/\epsilon \approx 1700$, which represents the path length multiplication factor or geometric gain of the empty cavity. Using the classical notion of finesse applied to our cavity and defined from the reflection losses $\mathcal{A}$ by $F = -2\pi/ln\mathcal{A}$, we obtained $F \approx 10500$ *(11)*. In addition, since the Lambertian reflectivity is due to multiple elastic light scattering events by the frozen disorder of the quartz powder structure of the wall, the reflection process is deterministic. However, the extreme complexity and the symmetries of diffuse reflections are such that any incident photon is reflected as an apparently 'random' photon, in the sense that its wavevector is distributed according to Lambert's law, and it has a phase and a polarization state that are uniformly distributed on $[0,2\pi]$ and on the Poincaré sphere respectively. These statistical properties of reflected photons suggest a strong analogy with the random geometry and polarization of the photon gas emitted by the walls of a blackbody cavity.

The optical field generated at any point $P$ inside the cavity (Fig.1B), in the far-field relative to the scattering walls, can therefore be considered as a random field obtained by superimposing a very large number of elementary plane waves independently taken from a unique statistical distribution. This unique probability distribution does not depend on the position $r$ of point $P$ and jointly aggregates three independent variables: a wavevector $k$ isotopically distributed on the sphere $|k| = k_0 = 2\pi/\lambda_0$, a uniform phase $\phi$ on $[0, 2\pi]$, and a polarization state represented by a complex vector $d$ uniformly distributed on the Poincaré sphere (Fig.1D). The total field can be described as highly composite random variable constructed as the sum of a random number of independently and identically distributed plane waves $E_\alpha(r,t)$. It writes

$$E(r,t,\mathcal{C}) = \sum_{\alpha=1}^{N_\mathcal{C}} a_\alpha e^{i\phi_\alpha} d_\alpha e^{ik_\alpha r - i\omega_\alpha t} \qquad [1]$$

where $N_\mathcal{C}$ is the number of elementary plane waves. $\mathcal{C}$ refers to one generic choice of the total random field, *i.e.* a microscopic configuration of the field. From the Huygens-Fresnel principle, we know that the optical field at $P$ is determined by the field on the cavity boundary, which comes with a very large but finite number of degrees of freedom. This sets a minimal value for $N_\mathcal{C}$ in equation 1. A lower estimate of $N_\mathcal{C}$ is given by the number of degrees of freedom associated with wavevectors, which typically amounts to $\Sigma_c/\lambda_0^2 \approx 10^{10}$ where $\Sigma_c$ is the surface area of cavity wall. Equation 1 exactly corresponds to what is known as the 3D random wave model *(12)*, which will be referred here to as 3D Berry field. It has been the focus of several theoretical numerical investigations but has never been used to formally describe an experimental realization of a 3D speckle field, to the best of our knowledge. Nor it has been used in the context of interferometry. In our experiments, the randomness and the high statistical symmetry of the optical field arises from the linear transformation of the input laser field by the extreme but deterministic complexity of multiple scattering by the frozen disorder of the wall structure. This highly complex boundary condition act as a feedback that leads to the interferometric properties of the 3D Berry field.

The phase $\phi_\alpha$ of each field component $E_\alpha$ is related to the phase of the laser input transformed by a random number $n_r(\alpha)$ of reflections alternated with random chords across the cavity with a random travel time $\tau_c$. This leads to the notion that the field in $P$ at time $t$ combines the phases of the laser input taken at a large number of earlier time points $t - \tau_\alpha$ (see Fig.1B), where the values of $\tau_\alpha$ are the propagation times, or mode lengths between the laser input point and the probe point $P$. The statistical distribution of these propagation times is assessed from the response of the cavity to a $120ps$ laser pulse (Fig.1C), which leads to the average photon transit or residence time $\overline{\tau_\alpha} \approx 207ns$, and the average photon path length $\overline{\Lambda} \approx 62m$. From the mean chord length theorem *(13,14)*, it is known that the mean chord length $\overline{\Lambda}_c$ is given by the ratio $\overline{\Lambda}_c = 4V_c/\Sigma_c$ of the cavity volume to the wall area, regardless of the shape. We obtain $\overline{\Lambda}_c \approx 3.5cm$, the average number of reflections $\overline{n}_r \approx 1700$, and finally the geometric gain mentioned above $G_w = \bar{n}_r$. These observations lead to two key remarks on the set of phases $\{\phi_\alpha\}_\mathcal{C}$. First, since the average chord length $\bar{\Lambda}$ is 8 orders of magnitude larger than the wavelength, the same is true for the standard deviation of the path length distribution, and this strongly supports our assumption that the phases $\phi_\alpha$ are uniformly distributed on $[0,2\pi]$. Second, the aforementioned coherence length $\Lambda_{coh}$ of the laser exceeds the average $\bar{\Lambda}_c$ by a factor 1.5. This feature was set by design, to ensure that practically any pair of fields $\{E_{\alpha_1}, E_{\alpha_2}\}$ would be mutually coherent, since 90% of the paths are longer than $\bar{\Lambda}_c$. Therefore, the 3D coherence matrix for any such pair of fields is a rank-two matrix. The coherent sum $E_{\alpha_1} + E_{\alpha_2}$ therefore has a 2D polarization, with a plane and a state of polarization that both depend on the polarization states $\{d_{\alpha_1}, d_{\alpha_2}\}$ and the phase difference $[\phi_{\alpha_1} - \phi_{\alpha_2}]$ *(15)*. The total field $E(r, t, \mathcal{C})$ is therefore fully polarized (Fig.1D), but its plane and state of polarization do strongly depend on the position $r$, with unique geometric and topological properties. In conclusion, in the absence of phase noise or any perturbation of the cavity, a unique microscopic configuration $\mathcal{C}$ is created and maintained, and the optical field $E(r, t, \mathcal{C})$ is a fully developed 3D spherical Gaussian speckle field. This field is associated with a position-dependent wavevector $k(r, \mathcal{C})$ and polarization state $d(r, \mathcal{C})$. Previous theoretical and numerical investigations of the 3D Berry field have revealed a rich set of entangled lines of optical singularities with unique geometric and topological properties *(16-20)*. In particular, the field lines associated with the field of wavevectors $k(r, \mathcal{C})$ are expected to exhibit strong curvature radii that can reach $\lambda$ *(18,19)*, and the optical field is highly structured with very complex geometry at the scale of $\lambda$. As a consequence, one each point of the aperture plane of the probe fiber, it is projected as a 2D circular Gaussian field, with an exponentially distributed intensity $I(r,C) = <E.E^*>_t (r,C)$ and a contrast close to unity (as detailed in the Supplementary Materials, section 2 "Speckle Statistics").

**Interferometric response of an unperturbed cavity**

Experimentally, the 3D speckle field is probed by introducing a single-mode fiber that equally splits the light toward two avalanche photodiodes, and a digital correlator (Fig.1A) which provides the intensity autocorrelation function $g^{(2)}(\tau)$ of the time-lag $\tau$. The normalized intensity autocorrelation function $|g^{(1)}|^2 = [g^{(2)}(\tau) - 1]/\beta$, is derived using the so-called coherence factor $\beta = g^{(2)}(0) - 1$ (Methods, section 3 "Data normalization"). In the unperturbed cavity, simply filled with air under atmospheric pressure at ambient temperature and thermal equilibrium with maximal environmental stability (Materials, section 3 "Environment control"), we reproducibly find a minute exponential decorrelation of amplitude $510^{-4}$ with a time-constant $\tau = 500\mu s$ (Fig.2A), due to the internal dynamics of the laser gain. While this decorrelation is not detectable on the usual linear [0-1] scale, a complete decorrelation is observed for very long time-lags, with a 50% decorrelation for a time-lag $\tau_{1/2}$ that increases with increasing acquisition time $T_{acq}$. Beyond $T_{acq} \geq 510^4 s$, a fixed value of $\tau_{1/2} \approx 3000s$ is reached, which reflects the decorrelation probably caused by the fluctuations of the central wavelength of the laser (Supplementary materials, section 3 "Sources of noise and digital correlation noises").

The unperturbed cavity thus provides a stable speckle field that decorrelates in a highly reproducible fashion because of known instrumental causes. Therefore, the normalized autocorrelation function of the unperturbed cavity can serve as the baseline of instrumental speckle fluctuations. To assess the interferometric response to various perturbations, we simply measure the excess decorrelation relative to the instrumental baseline, by computing the autocorrelation difference. The noise associated with this procedure, *i.e.* the root-mean square (*rms*) noise floor of our measurements, is a function of $\tau$ given by the standard deviation $\sigma_{|g^{(1)}|^2}(\tau)$ associated with the baseline (Fig.2B). This noise function which has multiple noise sources (Supplementary Materials, section 3 "Sources of noise and digital correlation noises") reflects the complex fourth-order statistics of the intensity and will be referred to as "baseline noise". It decreases for increased laser intensity and longer acquisition times. Since the primary signal, *i.e.* the decorrelation amplitude, cannot be larger than unity, the detection range has a ceiling fixed by the unity decorrelation. The dynamic range is therefore entirely determined by the minimal *rms* noise amplitude. The dynamic range reaches up to 6 (resp. 4) decades when the acquisition time is 3 (resp. 1) hours (Fig.2B).

**Interferometric response to picometric deformations**

To demonstrate how our 3D interferometer responds to minute deformations, the cavity was split in two halves, with two 1 mm thick piezo actuators inserted in between at a 120º angular distance to harmonically modulate the separation distance $\Delta l$ and hence the cavity shape and volume at 1 kHz (Fig.3A and Materials, section 4 "setup for piezo experiments"). As expected, the speckle intensity decorrelation exceeds the instrumental baseline by a difference that is modulated at the same frequency (Fig.3B), with a depth of modulation that grows linearly with the modulation amplitude of the piezo voltage and thickness (Fig.3C). In this set of experiments with a 3-hour acquisition time, the response coefficient for the decorrelation amplitude is $3.7 \times 10^{-5}$ pm$^{-1}$, the experimental noise is reached for a 2.7 pm piezo amplitude at a 1 kHz detection frequency, and the signal-to-noise ratio is $\approx 40$ for a 1 Å modulation amplitude. Using the experimental measurement of the aforementioned "baseline noise amplitude" (Fig.2B), we can express the power spectrum of the noise associated with the susceptibility to geometric deformations (Fig.3D). At 1 kHz, the noise amounts to $4 \times 10^{-3}$ pm$^2$ for a 3-hour acquisition time. This 3-hour acquisition time is the best compromise to mitigate the so-called "triangular averaging" noise due to the digitization process and the long-term laser wavelength noise, which respectively decrease and increase with the acquisition time (Supplementary Materials, section 3 "sources of noise and digital correlation noises" and Supplementary Materials, figure S7). Our data also suggest that the response is no longer linear for piezo amplitudes larger than 1 nm, because the decorrelation signal saturates when reaching unity. Although the modulation actuated by the piezo is in units of length, what is being probed are the variations of the 3D geometry (volume, surface and shape), which then boil down to variations of the geometric length invariant $\bar{\Lambda}_c$. The cavity deformation experiment primarily delivers the power spectrum of the variations of that length invariant.

The above results somehow contradict the observed non-ergodicity of the unperturbed cavity. By analogy with the sensitivity of a two-arm interferometer that is proportional to $|sin\Delta\phi|$ where $\Delta\phi$ is the phase difference between the arms, the interferometric response of the Berry field is different for each microscopic configuration $\mathcal{C}$. It is indeed determined by the specific contribution of the phase difference $\Delta\phi_{\alpha_1,\alpha_2}$ for each pair $\{E_{\alpha_1}, E_{\alpha_2}\}$ of elementary fields contained in $\mathcal{C}$. This is not what we observe, since the perturbed cavity indeed responds in a reproducible way. The fundamental reason is that any physical perturbation of the cavity shape can be considered to be microscopically irreversible. The speckle field thus undergoes a self-ergodization process caused by the perturbations of the cavity.

We can then estimate the sensitivity of our interferometer, by analogy with the Lorentzian finesse of a Fabry-Pérot, which is given by the ratio of the free spectral range over the full width at half maximum linewidth of the spectral line shape: $\mathcal{F} = \frac{\Delta\nu_{FSR}}{\Delta\nu}$. Since our interferometer has an equivalent free spectral range $\Delta\nu_{FSR} = \frac{c}{\Lambda_c}$ and a finesse $\mathcal{F} = 2\pi G_w$ (11), we can derive that our capacity to resolve a small change in wavelength goes as $\frac{\lambda^2}{2\pi G_w \Lambda_c}$. Independently of the Fabry-Pérot theory, the interferometric sensitivity of the cavity can also be directly derived from equation 1, leading to the same result. Practically, if we measure an intensity trace with $n_{ph}$ photons, with a frequency dependent signal-to-noise ratio $SNR(\tau)$ given by the noise power spectrum in Figure 2.B, we can write our detection limit as:

$$|\delta\lambda|_{min} = \frac{\lambda^2}{2\pi G_w \Lambda_c} SNR(\tau)/\sqrt{n_{ph}} = 1.1 \times 10^{-15} \times SNR(\tau)/\sqrt{n_{ph}} \quad [2]$$

which, for a typical number of photons $n_{ph} = 10^5$ photons and $SNR \geq 10$ gives $|\delta\lambda|_{min} = 3.5 \times 10^{-17} m$. To account for the observed picometer sensitivity, we consider that a *rms* variation of the path length distribution and a *rms* variation of the wavelength have conjugated effects on the phase, and hence have statistically similar effects: $\left|\delta\bar{\Lambda}_c\right|/\bar{\Lambda}_c = |\delta\lambda|/\lambda$, and it comes:

$$\left|\delta\bar{\Lambda}_c\right|_{min} \approx 1.6 \times 10^{-12} m \quad [3]$$

This simple theory leads to a 4-fold smaller sensitivity compared to experiments, and this underestimation is likely due to two fundamental difficulties. First, it is physically impossible to probe the actual 3D isotropic Berry field described by the theory, simply because the optical fiber truncates the field by casting a shadow that necessarily destroys its spherical symmetry. Second, the fiber does not even probe that truncated 3D field, but only its 2D isotropic projection on the fiber input plane. The fiber does not collect the theoretical 3D speckle intensity, but it probes the projected intensity of the truncated field, spatially integrated over the fiber aperture. Given the geometric complexity of the field lines of $k(r)$ and the field correlations at the scale of $\lambda$, the susceptibility to optical fluctuations is expectedly larger for the actual 2D "projected" speckle for the 3D Berry speckle intensity.

**Interferometric response to minute dielectric fluctuations**

To demonstrate how 3D stochastic interferometry probes optical fluctuations generated inside the cavity volume with unperturbed walls, we first introduced a dense and strongly scattering jammed emulsion made of calibrated oil-in-water droplets with a monodispersed radius of $459 \pm 15nm$. It has been extensively characterized for its optical and mechanical properties *(22)*. Given the very high volume-fraction used here (0.785), droplets are elastically jammed with small shape deformations, and they undergo sub-nanometer thermal motions that saturate at a root-mean-square (*rms*) amplitude of 900 pm which corresponds to the low-frequency plateau modulus. As a result, we find that the speckle intensity decorrelation is proportional to the time-lag $\tau$ for $10ns < \tau < 10\mu s$, and saturates at a plateau decorrelation of value of 6% (Fig.4A, purple line) with a cross-over time of $\tau = 70\mu s$ (Materials, section 5. Details on the experiments with intra-cavity samples). Using the known values of the *rms* displacement of droplets as a function of $\tau$ *(22)*, we find that the decorrelation scales linearly with the mean-square displacement (Figure 4A insert). Our results indicate that jammed droplets exhibit high-frequency Brownian fluctuations and demonstrate that we can measure *rms* motion amplitude up to 100 MHz with an unprecedented sensitivity: at the shortest time scale, a 25 pm *rms* motion amplitude is recovered.

In comparison, Figure 4A shows in blue a measurement on the very same sample and under the same acquisition conditions in a conventional Diffusing Wave Spectroscopy (DWS) in a transmission configuration *(23-25)*. While the DWS measurement also recovers the saturation plateau of longer time delays, it only provides a readable signal down to about $10^{-6}$s for a 314pm displacement. The noticeably less noisy signal from the cavity spans 8 frequency decades and is limited here by the acquisition rate of our digital correlator ($\tau_{sampling} = 12.5ns$). Our capacity to detect motions approximately one order of magnitude smaller than for DWS at frequencies two orders of magnitude higher for these samples clearly shows the amplification of multiply scattered light fluctuations. Our data also indicates that SNR≈2000 for the detection of 0.8 nm *rms* motions in the 0.1 to 100 kHz frequency range. This suggest that 8 pm motions could be detected with SNR=20 in that frequency range, with a dynamic range that can reach almost 4 decades. Keeping in mind that the decorrelation signal is proportional to the mean chord length of the sample, which amounts to 6.7 mm in these emulsion experiments, one could access the fluctuation dynamics of sub-millimeter sized scattering samples in miniature cavities with Angstrom level sensitivity.

To further demonstrate the interferometric response to intracavity optical fluctuations, we introduced 3 different suspensions of Mie scatterers ($10\mu m$ radius PMMA spheres, anisotropy coefficient g ≈ 0.957), with volume fractions such that the most (resp. least) concentrated sample is in the multiple (resp. single) scattering regime (Figure 4B). The speckle intensity decorrelation exhibits the expected monotonic increase, from the instrumental noise floor to unity decorrelation. The response combines a diffusive regime with a linear decorrelation with $\tau$, followed by a quadratic scaling dominated by sedimentation-induced motions (Fig.4B, purple line). Given the diffusion coefficient and the sedimentation velocity, the 1nm *rms* displacement expected for $\tau = 10\mu s$ is detected with SNR=30. In comparison, a measurement done on the same sample with a commercial Dynamic Light Scattering (DLS) (26) sizer instrument yields a measurement much noisier and more limited in dynamic range (Fig.4B, yellow line). Indeed, the signal is readable down to only $\approx 2 \times 10^{-4} s$ and is about 150 times smaller than the cavity signal. The cavity extends the frequency range by 2 order of magnitude and the dynamic range of decorrelation. By 1.5-2 decades. We also observe with our cavity that the decorrelation amplitude unexpectedly scales with the transport mean free path length as $1/\sqrt{l^*}$ (insert to Fig.4B). This observation points to need of a theory to better explain how the cavity amplifies the fluctuation amplitude of scattered light intensity compared to classical methods of DLS and DWS.

**Discussion**

In the present work, we have introduced the concept of 3D stochastic interferometry enabled by the feedback of a diffuse reflective cavity on a coherent and monochromatic photon gas. It detects minute variations of the "optical volume", as qualitatively defined by the geometry of the cavity, the dielectric tensor field inside. Thanks to the feedback on the 3D Berry field, the decorrelation function essentially detects all geometric and dielectric fluctuations through a 3D random paths integration process that is statistically isotropic and homogeneous, and fundamentally non-local. Indeed, the detection fiber statistically probes all geometric or dielectric perturbations regardless of where they occur. No information can be recovered on their spatial structure. More theoretical and experimental work is needed to quantitatively define this new concept of "optical volume", and especially to characterize quantitatively the statistical response of the random optical field to the variations of the dielectric tensor field. Beyond the preliminary theory sketched here, several other theoretical problems arise, such as the photon statistics of the intensity probed by the fiber *(27)*, the expectedly extreme degree of classical entanglement of the Berry field *(21)*, the consequences of its complex sub-wavelength structure and super-oscillations *(19)* on the interferometric properties, or a quantitative theory for the amplification of light scattering.

Practically, despite the simplicity of the design, our 3D interferometer provides a fluctuation spectrum over almost 10 frequency decades, with a dynamic range of 4 to 6 decades for the decorrelation signal and a picometric sensitivity. The resulting finesse F≈10500 competes favorably when compared to the best classical room-temperature Fabry-Perot interferometers designed with similar laser sources. Alternative designs can be envisioned in the frequency- and time-domain, with novel applications in the fields of light scattering and for ultrasensitive dielectric spectroscopy. In particular, thanks to the amplification of the fluctuation spectrum by the reflectivity gain $G_w$, the cavity can be used to measure dynamic light scattering from highly diluted or miniature samples *(35)* and quasi non-scattering objects such as proteins, thus providing the spectrum of internal modes of motion without labels or marker *(36)*. Cavity deformations can also be coupled to various external force fields (gravity, seismic motion or acoustic vibration, electromagnetic forces) for 3D picometer metrology and ultrasensitive detection of these force fields using a very simple design *(37)*.

**Materials**

1. **Construction and optical properties of quartz powder cavities**

Quartz cavities were made from a synthetic amorphous silica powder (ZANDOSIL 30, Heraeus) in a process first described by the group of Ed Fry *(10,28)*. Using transparent quartz molds ($1\,cm$ thick), the powder was mechanically compressed into a cylindrical shape, which was then baked and machined to obtain the desired cavity shape. The typical internal radius of such cylindrical cavities was $2.5\,cm$ while the height was $6\,cm$ and the thickness of the walls 3 cm (figure S8). To characterize the reflectivity of the cavity (albedo) and monitor its degradation induced by air humidity, it was measured using a ring-down method described by *(29)*. The response of the empty cavity to a laser pulse directly yields the reflection coefficient of the walls. Practically, a picosecond laser was used (PC-670M and PDL-800B, PicoQuant) with $120\,ps$ and $2\,nJ$ pulses at $670\,nm$, together with a Time Correlated Single Photon Counter (TCSPC PicoHarp 300 system, PicoQuant) with a $4\,ps$ time bin width, a dead time smaller that $95\,ns$, and 2 16-bits channels. The albedo of the quartz cavity was found to be $\rho = 0.9994 = 1 - 6 10^{-4}$, in agreement with the 0.99919 value found by Cone et al. *(10)* at $532\,nm$. The Lambertian reflectivity profile of the walls was confirmed through the measurement of the angular intensity profile using a goniometer equipped with a power-meter (Fig.S1). The expected cosine law was recovered.

### 2. Optical setup

The laser used for all interferometry measurements was a single frequency CW laser (Cobolt Flamenco 04-01 series), with a central wavelength $\lambda_0 = 659.6 \pm 0.3\,nm$ and an optical bandwidth $\Delta\nu \lesssim 1\,MHz$, *i.e.* a wavelength bandwidth $\Delta\lambda \lesssim 2.5\,fm$. This corresponds to a coherence time $\tau_{coh} = 1/\pi\Delta\nu \approx 0.3\,\mu s$ and a coherence length $\Lambda_{coh} = c/\pi\Delta\nu \approx 95\,m$. Typically, over 8 hours and when the temperature fluctuations do not exceed 2 °C, the central wavelength fluctuations saturate with a maximal standard deviation $\sigma_{\lambda_0} \lesssim 1\,pm$. The laser intensity is subjected to fluctuations with time that typically do not exceed 2% over 8 hours if temperature variations are less than 3 °C, and the relative *rms* intensity noise is less than 0.1% in the 20 Hz to 20 MHz frequency range. The laser is injected into the cavity with a multimode optical fiber (NA=0.22, FG050LGA, Thorlabs).

The speckle intensity was detected inside the cavity by inserting a single-mode fiber at $670 \pm 50\,nm$ with insertion loss less than 3.9 dB (TW630R5F1, Thorlabs). The fiber was split in two branches connected to two photon counting avalanche photodiodes (SPCM-AQRH, Excelitas) with the following characteristics: diameter 180 µm, 65% efficiency at 650 nm, 10 ns output pulse width, 22 ns dead time and 1.4% after-pulsing probability. The photodiodes were connected to a two-channel correlator (LS Instruments) operated in the 16/8 multi-tau correlation scheme with the following features: lag times spanning form 12.5 ns to 3436 s, 322 channels, a 54976 s lag time range, and a maximum count rate of $2 \cdot 10^7$ counts per second over 52 ms integration time intervals.

    3. **Environmental control**

All experiments were performed in an isolated acrylic box, covered with foam panels for thermal and acoustic isolation, and enclosed in a Faraday cage for precautionary measure against electro-magnetic noise. The setup was placed on an active isolation optic breadboard (DVIA-T, Daeil Systems) on top of a compressed-air passive isolation optical table. Hygrometry and temperature inside the box were continuously monitored, as well as the temperature of the laser which was passively cooled. All measurements were made after waiting a precautionary 30 min warm-up time of the cavity to ensure a steady-state behavior was reached. The temperature inside the cavity reached maximum fluctuations of about $\pm 0.1\,°C$ and did not drift from the timescale if minutes to days.

    4. **Setup for piezo experiments**

Two piezoelectric ceramics with perovskite structures (ABO$_3$) ((1-x)Ba(Zr$_{0.2}$Ti$_{0.8}$)O$_3$-x(Ba$_{0.7}$Ca$_{0.3}$)TiO$_3$ (x=0.5), Curie temperature = 100°C) were used as piezo actuators. These actuators had a thickness of $1 mm$ and a surface of 0.8 m$^2$. They were inserted in between the two halves of the cavity and their thickness was harmonically modulated at $1 kHz$ with a range of voltage amplitudes using a function generator. Using piezo-response force microscopy, the piezoelectric coefficient was measured at about $d_{33} = 300 \pm 15 pC/N$ and the ferroelectric domains were visualized with the same technique as well as polarization/strain graphs (Fig.S3).

### 5. Details on the experiments with intra-cavity samples

For experiments conducted with samples inside the cavity (Main, Fig.4), the setup can be characterized by the geometric parameters shown on Fig.S2:

1. the volumes (in cm$^3$) are $V_c = 118$, $V_{s1} = 0.785$ and $V_{s2} = 19.6$ for the cavity, the jammed emulsion and suspended microsphere samples respectively.
2. the surface areas (in cm$^2$) are respectively $\Sigma_c = 134$, $\Sigma_{s1} = 4.71$ and $\Sigma_{s2} = 55$.
3. the resulting mean chord (in cm) are respectively $\bar{\Lambda}_c = 3.53$, $\bar{\Lambda}_{s1} 0.67$ and $\bar{\Lambda}_{s2} = 1.43$ given by the mean-chord length formula $\bar{\Lambda} = 4V/\Sigma$.

The first series of samples (main text, Fig.4a) consists in a jammed emulsion of monodisperse polydimethyl siloxane (PDMS) oil-in-water prepared and characterized by Kim et al. *(22)* (kind gift of Prof. Thomas Mason, UCLA). The average droplet radius is 459±15 nm with refractive index 1.401±0.001 at 660 nm, and their volume fraction is 0.738. The resulting transport mean free path is $l_{s1}^* = 0.25\,mm$. Given the size of emulsion samples, the mean chord length $\bar{\Lambda}_{s1} = 6.7\,mm$ exceeds $l_{s1}^*$ by a factor 27. Light is thus multiply scattered by the jammed emulsion, and the photons follow a Brownian excursion through the sample with typically 700 scattering events. In addition, because of the path invariance in scattering media [5], the average length of Brownian paths before exiting the sample volume equals $\bar{\Lambda}_{s1}$ regardless of $l_{s1}^*$. Given the estimated absorption length $l_{s1}^{(a)} = 10^4\,mm$, the absorption probability along the Brownian path is given by $l_{s1}^*/l_{s1}^{(a)} \approx 1/4000$ and can be safely neglected. However, the effect on the albedo is obtained by multiplying this very small absorption probability by the probability of a random cavity chord to intersect the volume sample, which is given by the view factor of the sample from the cavity. This factor can be estimated by the surface ratio (≈1/30), and the estimated contribution to the loss typically amounts $10^{-5}$. Practically, we found a larger reflection loss, probably due to the sample container absorption. The effective albedo drops from 99.94% to 99.8% and the cavity gain is reduced from 1700 to about 500, thus reducing the overall amplifying effect of the cavity.

The second series of samples (main text, Fig.4b) consists in R=10 µm radius polymethyl-methacrylate (PMMA) particles in water (Bangs Laboratories), with refractive index 1.488±0.001 at 660 nm. Given the particle size and the refractive index difference with water (0.156±0.001), the scattering anisotropy coefficient is g=0.9571, and the root-mean light deviation angle is $\sqrt{\langle\delta\theta^2\rangle} \approx 11°$. With a density of 1.19 $kg/m^3$ at 20 °C, the density difference with water is $\Delta\rho = 0.192 \pm 0.001\,kg/m^3$. Knowing the viscosity of water $\eta = 10^{-3}\,Pa.s$ and the gravity $g$, the resulting sedimentation velocity given by the Stokes Law is $V_{sed} = \frac{2R^2 g \Delta\rho}{9\eta} =$ 42$\mu m/s$ = 42$pm/\mu s$. From the Stokes-Einstein equation, the diffusion coefficient is $D = k_B T/6\pi\eta R = 2.110^4\,pm^2/\mu s$ (with the Boltzmann constant $k_B$ and the temperature $T$). The motion of these particles in water crosses over from a short-time regime dominated by thermally activated Brownian diffusion to a long-time regime dominated by gravity-driven sedimentation. The cross-over time given by $\langle\delta r^2(\tau)\rangle_t = 6D\tau = V_{sed}^2 \tau^2$ is $\tau_{cross} = \frac{6D}{V_{sed}^2} = 71\mu s$.

The volume fractions used were 0.1% (transport mean free path $l^* = 0.7\,mm$), 0.01% ($l^* = 7\,mm$) and 0.001% ($l^* = 70\,mm$). The sample volume was $V_{s2} = 19.6\,cm^3$, its surface $\Sigma_{s2} = 55\,cm^2$, and its mean chords length was $\bar{\Lambda}_{s2} = 1.43\,cm$. Because the values of the ratio $\bar{\Lambda}_{s2}/l^*$ indicating the scattering multiplicity are respectively 20, 2, and 0.2 for decreasing volume fractions, the first sample produces multiple scattering while the third one is in the single scattering regime. Even for the largest volume fraction, the probability of absorption through the sample is far below the wall reflection loss coefficient $\epsilon$.

**Methods**

1. **Cavity geometry - statistics - photon budget**

    1.1  **Statistics of reflections & intracavity radiance**

Thanks to the high diffuse reflectance of cavity walls, the optical intensity $P_0$ injected into the cavity ends up being multiplied and uniformly reflected by the cavity walls. The reflection efficiency can be considered at the quantum level by a photon survival probability that corresponds to the Albedo $\mathcal{A}_w$. At steady-state, uniform reflection losses exactly balance the intensity $P_0$, and the walls receive a uniform irradiance $\mathcal{I}_{rc}$ over a total surface area $\Sigma_c$, such that $P_0 = (1 - \mathcal{A}_w)\Sigma_c \mathcal{I}_{rc}$. Consequently, the wall radiance originates from the irradiance as $\mathcal{R}_c = \mathcal{A}_w \mathcal{I}_{rc}/\pi$. The number $n_r$ of reflections follows a geometric distribution with probability $p(n_r) = \mathcal{A}_w^{n_r}(1 - \mathcal{A}_w)$ for $0 \leq n_r$, and its mean gives the cavity gain $G_w$ as:

$$\overline{n_r} = G_w = \mathcal{A}_w/(1 - \mathcal{A}_w) \quad [4]$$

The total intensity reflected by the wall surface is therefore $G_w P_0$, and the wall radiance is:

$$\mathcal{R}_c = G_w P_0/\pi \Sigma_c \quad [5]$$

Practically, with the relatively large gain we obtained, namely $G_w \approx 1700$, and using a typical input intensity of $P_0 = 100\,mW \approx 3.3 \times 10^{17} ph.s^{-1}$ injected into a cylindrical cavity with a height $h_c = 6 cm$ and a diameter $\phi_c = 5 cm$, $\Sigma c = 134 cm^2$ and the radiance is $\mathcal{R}_c = 5 \times 10^{21} ph.s^{-1}.m^{-2}.sr^{-1}$. The losses of energy at the cavity walls are small: $G_w \approx 6 \times 10^{-4}$, that optical energy is very slightly imbalanced in their vicinity. This is not true at the point of laser injection, where the local radiance exceeds the average wall radiance by a factor $(1 - \mathcal{A}_w)\Sigma_c/\Sigma_{laser}$, where $\Sigma_{laser}$ is the effective cross-section area of the laser input. This heterogeneity of the wall radiance can be mitigated by installing a simple baffle screen.

As a result, if we virtually introduce a small planar two-sided object with two opposite surfaces $\pm\vec{\delta}_\sigma$ at a random point inside the cavity, we can safely assume that both surfaces will be receiving the same irradiance $\mathcal{R}_c$ but with opposed orientations. We can also safely assume that the same irradiance will impact the entrance surface of any light detection device introduced inside the cavity, such as the surface of a small camera, or the input cross-section of a fiber introduced to collect light as described in the Materials and Methods, section 2. Optical setup. However, the light collected by such objects is removed from the intracavity power density and this leads to a reduction of the effective Albedo. This is why the effective Albedo must be calibrated for each cavity configuration, and this precaution obviously includes the situations where a possibly absorbing sample is introduced into the cavity. Under this provision, the cavity gain $G_c$ and the cavity radiance are correctly given by from the effective Albedo $\mathcal{A}$ determined from the pulse response by:

$$\mathcal{R}_c = G_c P_0 / \pi \Sigma_c \qquad [6]$$

$$\bar{n}_r = G_c = \mathcal{A}/(1-\mathcal{A}) \quad [7]$$

### 1.2 Statistics of intracavity chord length & photon path length

The pulse response function (Fig.1 in the main text) indicates the residence time of an optical pulse energy inside the cavity, or the typical time a photon survives before it is absorbed by the walls. A simplistic picture can be considered, in which a random photon path is made of a succession of a random number $n_r$ of reflections, and $n_c = n_r + 1$ free-space chords through the cavity. These chords can be represented by the random variable $\Lambda_c$, and the well-known mean-chord-length property applies (13,14): for any finite volume $V$ enclosed by a surface $\Sigma$ and filled with any perfectly scattering medium, be it homogeneous and isotropic or not, and regardless of the geometry of the cavity or the medium, a particle entering the volume with a uniform and isotropic incidence will follow a random trajectory inside the volume with a mean given by the very simple geometric invariant $4V/\Sigma$. For our cavities, this property implies that the mean of free-space chords with cavity volume $V_c$ is:

$$\bar{\Lambda}_c = 4\frac{V_c}{\Sigma_c} \quad [8]$$

The free-space path length variable $\Lambda$ is therefore the product of the two independent random variables $n_c$ and $\Lambda_c$, which averages to $\overline{n_c \Lambda_c} = \overline{n_c}\,\overline{\Lambda_c} = 4G_c V_c/\Sigma_c$.

Besides the free-space contribution to the photon path, the diffusive random walk inside the walls needed for the reflection process corresponds to an additional random variable $\tau_r$ that reflects the random reflection time. Its mean has been measured for visible light as a typical delay of $\bar{\tau}_r \approx 5ps$. As a consequence, the total reflection time averages as $\overline{n_r \tau_r} = \overline{n_r}\overline{\tau_r} \approx G_c \overline{\tau_r}$. The total photon path length can be represented as a composite random variable

$$\Lambda = n_c \Lambda_c + n_r c \tau_r \quad [9]$$

with the variance $\sigma_\Lambda^2$ and the average $\bar{\Lambda}$ given by:

$$\bar{\Lambda} = G_c \left[ \bar{\Lambda}_c + c\bar{\tau}_r \right] \quad [10]$$

Practically, in the conditions of our setup, with cylindrical cavities with $h_c = 6cm$ and $\phi_c = 5cm$, it comes:

$$\bar{\Lambda}_c = h_c / \left( \frac{1}{2} + \frac{h_c}{\phi_c} \right) \quad [11]$$

and $\bar{\Lambda}_c = 3.5cm$, while $c\tau_r = 1.5mm$. The mean total photon path therefore corresponds to a length of $\bar{\Lambda} \approx 62m$ and a mean residence time $\bar{\tau} = \bar{\Lambda}/c \approx 207ns$. Importantly, the coherence length of the laser exceeds the average photon pathlength by a factor $\Lambda_{coh}/\bar{\Lambda} \approx 1.5$. Less than 10% of the photon path are longer than the coherence length. This leads us to safely assume that most pairs of modes are mutually coherent, and the optical field is practically coherent across the cavity, thus enabling multiphoton interferences.

## 1.3 Photon density

The light inside the cavity can be considered as an isotropic and monochromatic photon gas, with a homogeneous density $\rho_{ph}$. This density can be computed from the photon energy, $h_c/\lambda_0$, the photon injection rate $P_{0,ph} = P_0 \lambda_0 / h_c$, and the mean residence time $\bar{\tau}$ described above. It comes $\rho_{ph} = P_{0,ph} \bar{\tau}/V_c$, or $\rho_{ph} = 4\pi \mathcal{R}_{c,ph}/c.G_c$. With $P_0 = 100mW$ as considered above, the photon injection rate is $P_{0,ph} \approx 3.3 \times 10^{17} ph.s^{-1}$, the photon density is $\rho_{ph} \approx 1.3 \times 10^9 ph.cm^{-3}$ and the photonic radiance is $\mathcal{R}_{c,ph} \approx 5 \times 10^{21} ph.s^{-1}.m^{-2}.sr^{-1}$.

## 1.4 Capture of the speckle intensity by a monomode fiber

Using the photon gas picture, we can estimate the rate at which photons can be collected from the cavity with the single-mode fibers used in our experiments. Knowing the surface of the collect optical fiber, we can use the concept or irradiance to calculate the number of detected photons. Here, the irradiance is $\mathcal{E}_{c,ph} = \pi.\mathcal{R}_{c,ph}$ so that the flux through an optical fiber with a core of cross-section $\sigma_{det} = 19\mu m^2$ (diameter $5\mu m$) would be about $1.2 \times 10^{12} ph.s^{-1}$ or $0.5 \times 10^{-6} W$. Taking the fiber losses into account (excess loss (0.3dB), insertion loss (4.2dB), optical return loss (60dB)) the flux detected is about $1.7 \times 10^{-10} mW$ or $5.6 \times 10^5 ph.s^{-1}$ which agrees with the typical experimental detection rates of $10^4$ to $10^5 ph.s^{-1}$. We should also stress here that the measurement fundamentally perturbs the speckle pattern as the fiber destroys the spherical symmetry of the field statistics.

## 2. Data acquisition

The purpose of the acquisition and analysis is to measure the temporal fluctuations of the speckle intensity, from the autocorrelation of the intensity entering the detection fiber inside the cavity. The normalized autocorrelation function of the intensity is mathematically defined for the time delay $\tau$ by

$$g_I(\tau) = \frac{\langle I(t)\rangle\langle I(t+\tau)\rangle_t}{\langle I^2\rangle_t} \qquad [12]$$

but the usual autocorrelation functions $g^{(1)}(\tau)$ and $g^{(2)}(\tau)$ used in experiments are not normalized the same way. The field autocorrelation function reads:

$$g^{(1)}(\tau) = \frac{\langle E(t+\tau)E(t)^*\rangle_t}{\langle E(t)E(t)^*\rangle_t} \qquad [13]$$

while the intensity autocorrelation function is defined by

$$g^{(2)}(\tau) = \frac{\langle I(t+\tau)I(t)\rangle_t}{\langle I(t)\rangle_t^2} = \gamma g_I(\tau) \qquad [14]$$

where the factor $\gamma = \langle I^2\rangle_t/\langle I\rangle_t^2$ reflects the statistics of the intensity fluctuations. These two latter correlation functions are connected by the Siegert relation $[g^{(2)}(\tau) - 1] = \beta|g^{(1)}(\tau)|^2$ used to define the so-called coherence factor $\beta$, which is usually determined by the instrumental design. Since $g^{(1)} = g_I = 1$ for $\tau = 0$, we obtain $\gamma = 1 + \beta$. The coherence factor is an important parameter for dynamic light scattering experiments, which reflects the statistics of the intensity fluctuations due to the degree of coherence of the optical field over the area where the intensity is probed. Practically, since the autocorrelator delivers the values of $\beta |g^{(1)}(\tau)|^2$, the normalization procedure described below requires that each data be normalized by the particular value of $\beta$ it contains.

Technically, the intensity autocorrelation function is delivered in real time by the correlator for a set of times delays $\tau$ according to a correlation scheme classically called 16/8 multi-tau. These time delays are logarithmically separated, as multiples of $\tau_0$, between a minimal value $\tau_0 = 12.5 ns$ that corresponds to the sampling time of the photodiodes, and a maximal value of 3436 s. The primary data handled by the correlator is a time series of binary counts $n(t_i)$ sampled with the minimal delay $t_{i+1} - t_i = \tau_0$. For each time delay $\tau$, the instruments compute the running average count rate defined by

$$n_0 = \frac{1}{N(\tau)} \sum_1^{N(\tau)} n(t_i) \qquad [15]$$

and the running autocorrelation value defined by

$$G(\tau) = \frac{1}{N(\tau)} \sum_1^{N(\tau)} n(t_i) n(t_i - \tau) \qquad [16]$$

where $N(\tau)$ is the total number of samples for the time delay $\tau$. From $n_0$ and $G(\tau)$, the correlator updates in real time the function

$$g^{(2)}(\tau) = \frac{G(\tau)-n_0^2}{n_0^2} \qquad [17]$$

and delivers the values of the function $g^{(2)}(\tau) - 1 = \beta |g^{(1)}(\tau)|^2$.

In order to avoid two major artifacts caused by the binary detection of photons by APDs, namely the issues of dead-time and afterpulse, we systematically used a split fiber and performed the cross-correlation between the photon streams detected by the two photodiodes. In the cross-correlation algorithm, the squared running average $n_0^2$ is simply replaced by the product $n_0^{(det\#1)} \times n_0^{(det\#2)}$ of the running averages, and the correlation $G(\tau)$ is computed from the cross products $n^{(det\#1)}(t_i)n^{(det\#2)}(t_i - \tau)$ instead of $n(t_i)n(t_i - \tau)$.

### 3. Data normalization

As mentioned above, virtually all correlators primarily deliver the function $g^{(2)}(\tau) - 1 = \beta |g^{(1)}(\tau)|^2$ which decreases from $\beta$ to 0 and is usually represented in a semilog-x or linear scale, in the usual data handling procedure, since $|g^{(1)}(\tau)| = 1$, $\beta$ is assessed from the $\tau = 0$ intercept of $\beta |g^{(1)}(\tau)|^2$, and then used to produce the normalized autocorrelation function $|g^{(1)}(\tau)|^2$. Much like the factor $\gamma$ defined by equation 14, the coherence factor $\beta$ takes values between 1 and 0, depending on the degree of coherence of the optical field on the surface over which the intensity is integrated. In principle, $\beta$ only depends on the area on which the intensity is sampled, it thus takes a constant value for a fixed light collection geometry.

## 4. Data representation

This section describes the graphical representations introduced in this paper (see Fig.S4) to:

a. evidence very small decorrelation amplitudes more clearly

b. fully exploit the large dynamic range of our cavity interferometer (up to 6 decades)

c. exhibit signal-to-noise ratio by displaying the noise together with the signal.

Autocorrelation data is classically represented by plotting the normalized autocorrelation function

$$\frac{g^{(2)}(\tau)-1}{\beta} = |g^{(1)}(\tau)|^2 \qquad [18]$$

on a semilog-x graph. It shows the decay on a linear vertical scale from 1 to 0 with a rather limited dynamic range. To exhibit small decorrelation amplitudes with a large dynamic range, we replace the linear vertical scale by a logarithmic one, on which we plot the decorrelation function defined as $1 - |g^{(1)}(\tau)|^2$. However, experiments done with the empty unperturbed cavity systematically exhibit an autocorrelation decay due to a 2 kHz gain relaxation and finite sample noise (see next section). Therefore, decorrelation measurements only make sense when compared to the 'reference' decorrelation function measured with the unperturbed cavity. As a consequence, the decorrelation $|g^{(1)}(\tau)|^2_{measure}$ obtained in a given experiment should not be compared to the function $|g^{(1)}(\tau)|^2 \equiv 1$, but to the decorrelation $|g^{(1)}(\tau)|^2_{ref}$ obtained from reference experiments conducted with the unperturbed cavity. The relevant signal is therefore the excess of decorrelation given by the difference:

$$\mathcal{E}(\tau) = [1 - |g^{(1)}(\tau)|^2_{measure}] - [1 - |g^{(1)}(\tau)|^2_{ref}] \qquad [19]$$

Practically, the data is processed as follows (see Fig.S4): for all experimental conditions (defined by the laser power and the acquisition time), a set of reference decorrelation functions $\beta|g^{(1)}(\tau)|^2$ is acquired with the unperturbed cavity (Fig.S4(a)). Since each function comes with a particular value of $\beta$ as explained in the previous section, they need to be individually normalized by their own value of $\beta$. From the set of normalized reference functions $|g^{(1)}(\tau)|^2$ (see Fig.S4(b)), we can derive an average reference for $|g^{(1)}(\tau)|^2_{ref}$ (Fig.S4(c)) and measure the statistical quality of that average reference by estimating the standard deviation $\sigma_{|g^{(1)}|^2_{ref}}(\tau)$. To measure the cavity response to some perturbation, the same scheme is applied, with the acquisition of multiple samples of $\beta|g^{(1)}(\tau)|^2$ (Fig.S4(a)), the individual normalized by their $\beta$ values (Fig.S4(b)), and the computation of the average normalized response $|g^{(1)}(\tau)|^2_{measure}$ (Fig.S4(c)). The temporal structures of the reference decorrelation and the perturbation decorrelation (respectively blue and red curves on Fig.S4(c)) can be more easily inspected by plotting their complement to 1, i.e. $1 - |g^{(1)}(\tau)|^2_{ref}$ and $1 - |g^{(1)}(\tau)|^2_{measure}$ (Fig.S4(d)), but the meaningful signal is their difference $\mathcal{E}(\tau)$, which measures the decorrelation caused by the perturbation and defined by equation 11. Finally, $\mathcal{E}(\tau)$ is plotted with a logarithmic vertical scale, together with noise floor given by the experimental standard deviation introduced above as $\sigma_{|g^{(1)}|^2_{ref}}(\tau)$ (Fig.S4(e)). If the measured perturbation has no effect, the decorrelation excess function $\mathcal{E}(\tau)$ should be exactly equal to the noise floor function $\sigma_{|g^{(1)}|^2_{ref}}(\tau)$ multiplied by $\sqrt{2}$. As a consequence, Fig.S4(d) directly indicates the signal-to-noise ratio ($SNR$) up to a factor $\sqrt{2}$, by the logarithmic 'distance' between the signal $\mathcal{E}(\tau)$ and the noise floor (respectively the black and the green curve).

**Acknowledgments**

The authors are grateful to Sir Michael Berry, Profs. John King, Joerg Enderlein, Philippe Poulin, Shankar Ghosh and Rodney Ruoff for comments and critical reading of the manuscript. Profs. Ed Fry and John Mason are acknowledged for their initial help with quartz cavities, Prof. Tom Mason and Prof. Jo Wook for providing us with jammed emulsion samples and piezo actuators respectively. Mr Kim Hwanhee is acknowledged for some data acquisition in figures 3 and S8. This work (G.G. & F.A.) was supported by the taxpayers of South Korea through the Institute for Basic Science, Project Code IBS-R020-D1. G.G. was also supported by the National Research Foundation of Korea (NRF) grant funded by the Korea government (MSIT) (No. 2021R1A3B1071354 to Tae-Young Yoon). M.F. is supported by Simons Foundation Grant No. 601944, MF. F.A. acknowledges the remote support of the Laplacians' team.

**Author contributions**

F.A directed the project, G.G conducted the research and experimental measurements, F.A., M.F. and G.G developed the theory and F.A, M.F. and G.G. wrote the manuscript.

**Competing Interests**

The authors declare that they have no competing interests.

**Data and materials availability**

All data needed to evaluate the conclusions in the paper are present in the paper and/or the Supplementary Materials.

**Supplementary Materials:**

Supplementary Figures S1-S8

Supplementary text

**References and Notes:**

1. Fizeau, MH. On the hypotheses relating to the luminous æther, and an experiment which appears to demonstrate that the motion of bodies alters the velociety with which light propagates itself in their interior. *The London, Edinburgh, and Dublin Philosophical Magazine and Journal of Science* **2.14,** p. 568-573 (1851)

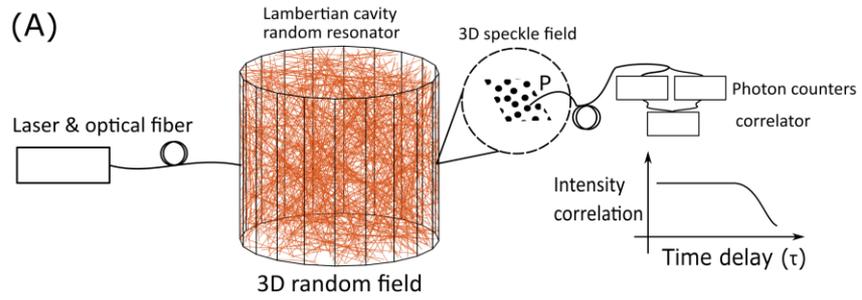

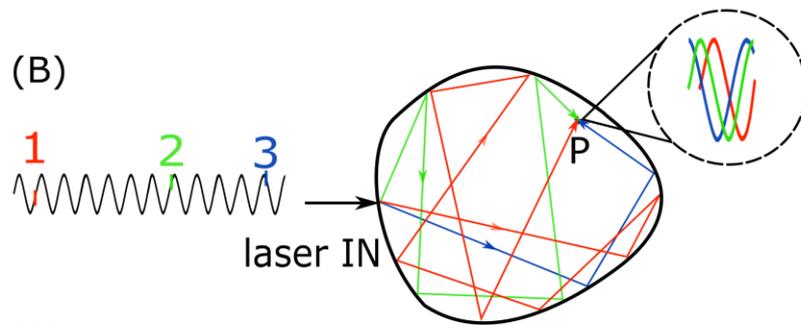

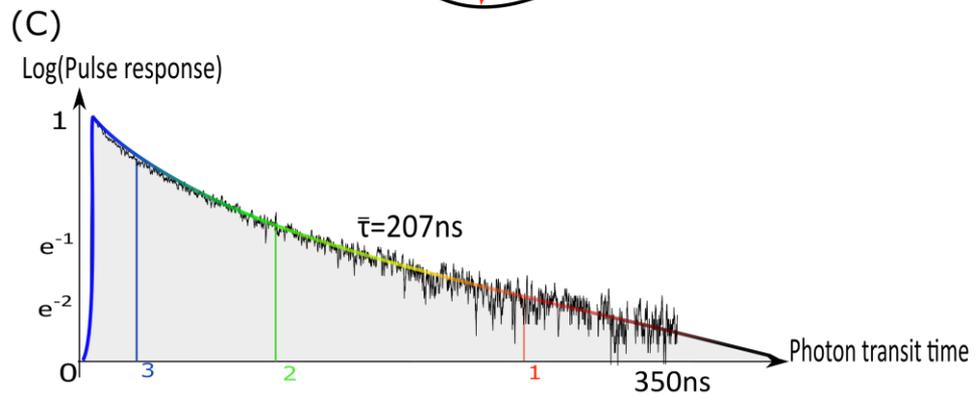

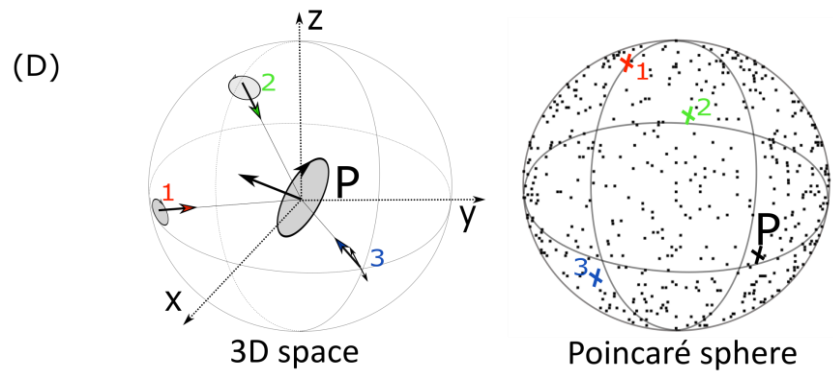

**Fig. 1.** 3D stochastic interferometer (A) Experimental setup: a single-frequency laser is injected by an optical fiber into a high reflectivity Lambertian cavity. Regardless of the cavity shape, the resulting 3D random field is fully coherent and generates a fully developped 3D speckle pattern which is probed at point P by a single mode fiber. The fiber is split, and the intensity is detected by 2 photon counters. The intensity cross-correlation is then calculated by a digital correlator. (B) schematic representation of 3 generic path photons or modes, as a succession of diffuse reflections, between the laser input and a probe point P. (C) Intensity response function of a compressed quartz powder cavity to a 120 ps laser pulse at 670 nm represented in a y-log scale. (D) Left: Coherent superposition of 3 plane waves at point P, with their respective complex polarization. (D) Right: Poincaré sphere representing the uniformly and randomly distributed polarizations of the plane waves interfering at point P. In particular, the polarization states of the aforementioned 3 plane waves are represented, as well as the 2D polarization state of the wave generated at point P.

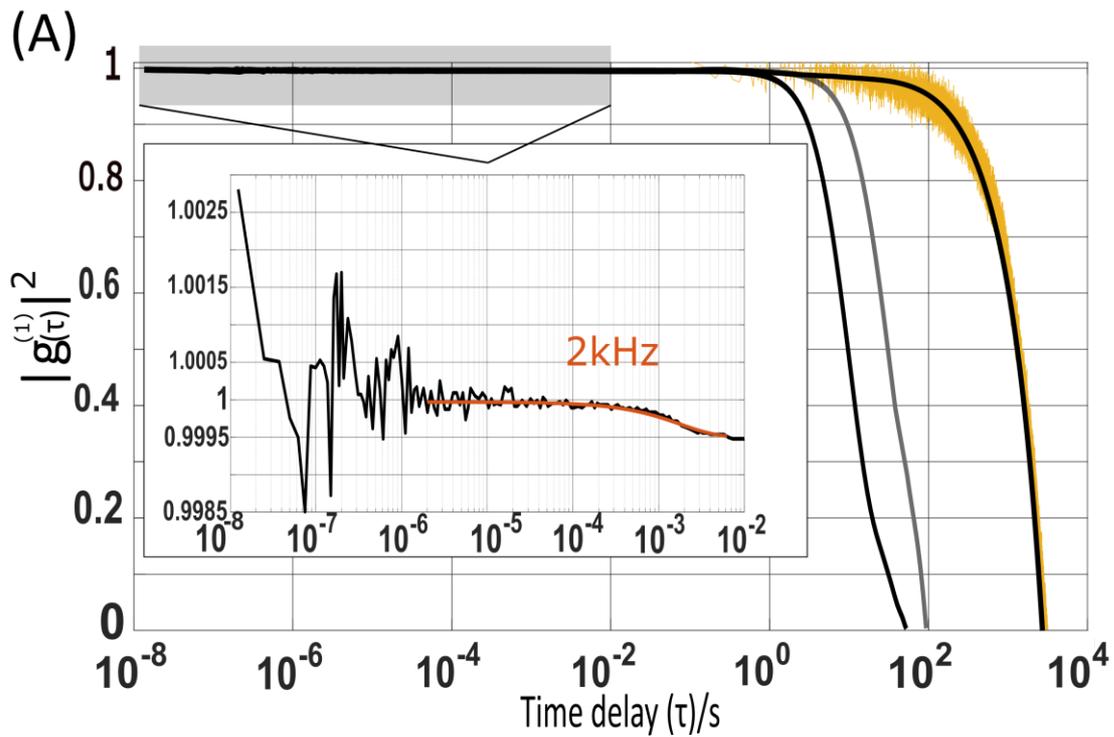

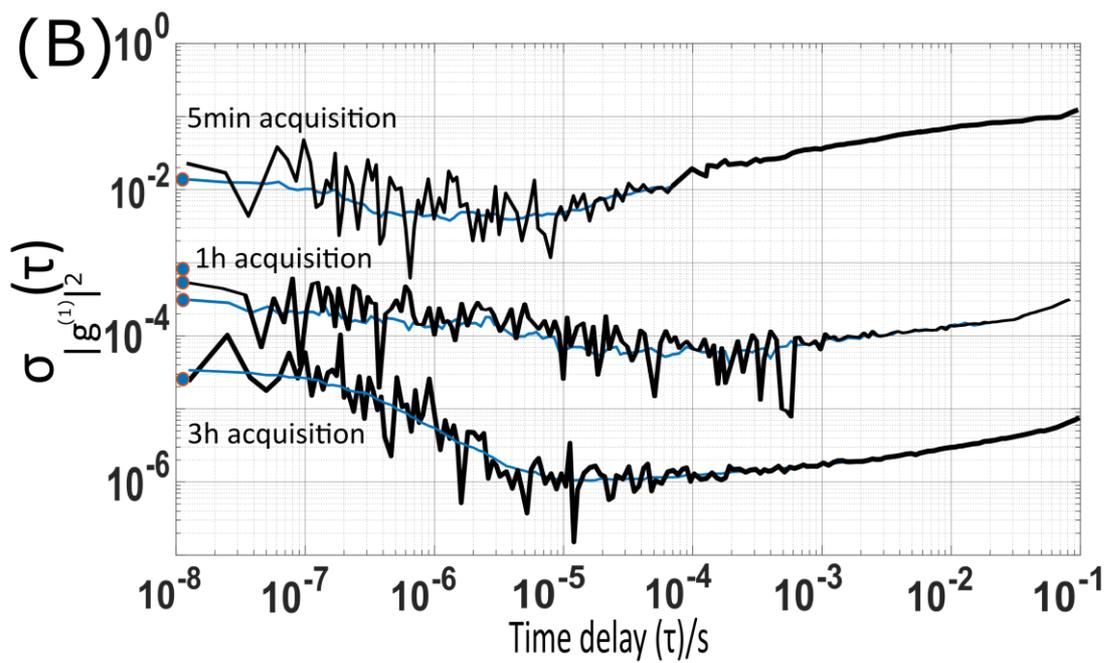

**Fig. 2.** Properties of the speckle intensity autocorrelation in the unperturbed quartz cavity. The laser input power is 300 mW if not stated otherwise and the light is collected with a monomode fiber. (A) Normalized intensity auto-correlation $[g^{(2)}(\tau) - 1]/\beta$ as a function of the time delay $\tau$. From left to right, acquisition times $T_{acq}$ in seconds are 2.5 10³, 3.6 10³, and 65 10³. Each dark curve is the average of 3 repetitions, and the yellow curve represents one of those repetitions. The insert magnifies the initial decorrelation for $T_{acq}$= 3.6 10³ s fitted by an exponential decay at 2 kHz (red line). (B) Noise (standard deviation) of the intensity autocorrelation function (black lines) measured for different acquisition times. The blue lines are moving averages of the raw data and the blue dots at the shortest time delay correspond to $T_{acq}$ in minutes of 5, 60 and 180 from top to bottom.

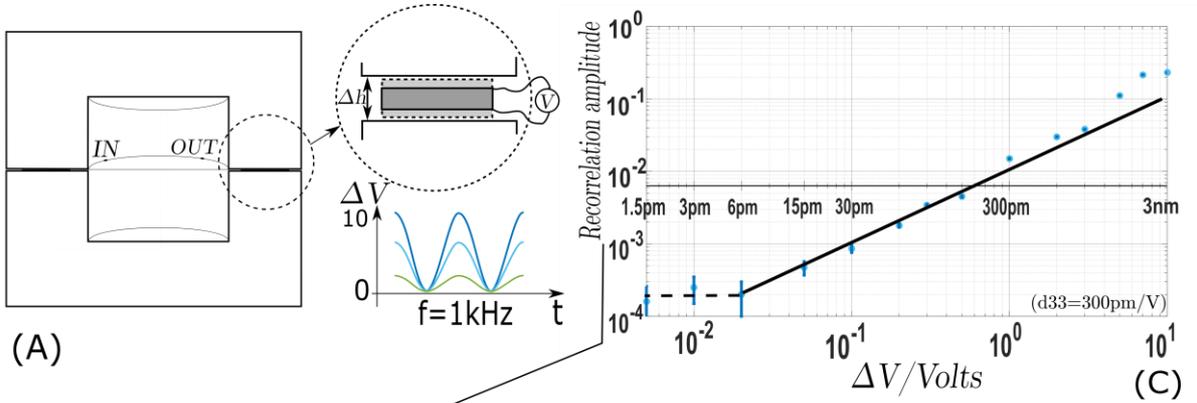
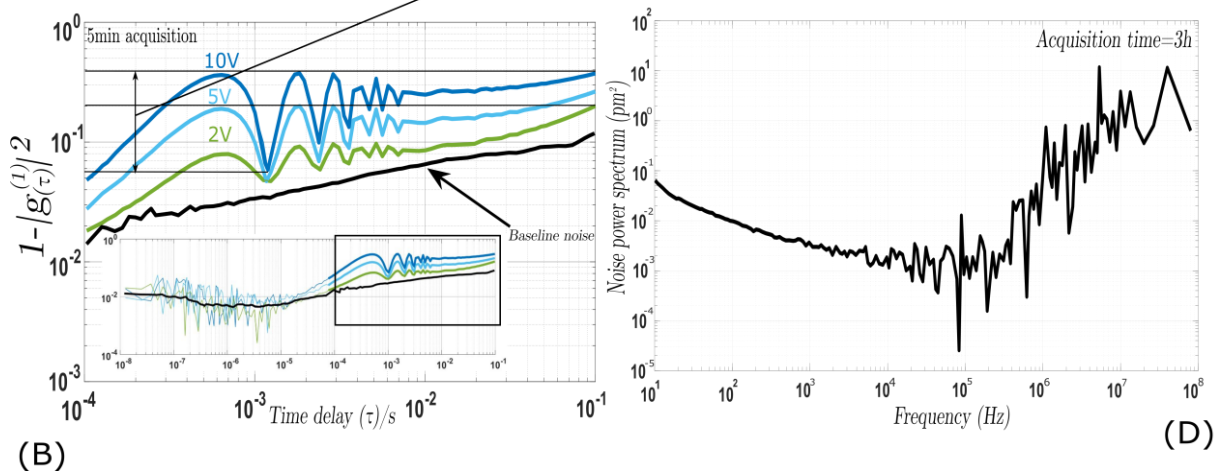

**Fig. 3.** Interferometer response to harmonic cavity shape deformations. (A) Home-made cylindrical quartz cavity (height $h_c$=6 cm, diameter $\varphi_c$=5 cm, wall thickness 3 cm). Two voltage driven piezo actuators (thickness 1 mm, placed at a 120° angle) are inserted between the two halves of the cavity. A sinusoidal 1 kHz voltage is fed to the piezo actuators to deform the cavity. (B) Intensity decorrelation of the speckle intensity $[1 - |g^{(1)}(\tau)|^2]$ as a function of time delay $\tau$ for voltage amplitudes of 10, 5 and 2 V. The noise baseline is the standard deviation $\sigma_{|g^{(1)}|^2}(\tau)$ of the auto-correlation function for the unperturbed cavity (black line). The acquisition time was 5 min with 300 mW laser input power, and experiments were averaged over 10 repetitions. (C) Recorrelation amplitude obtained from (B) as a function of the piezo voltage and actuation amplitude. The acquisition time was 5 min for voltages above 1 V and 30 min below (D) Noise power spectrum of the system expressed in pm$^2$ as a function of frequency, for a 3-hour acquisition time.

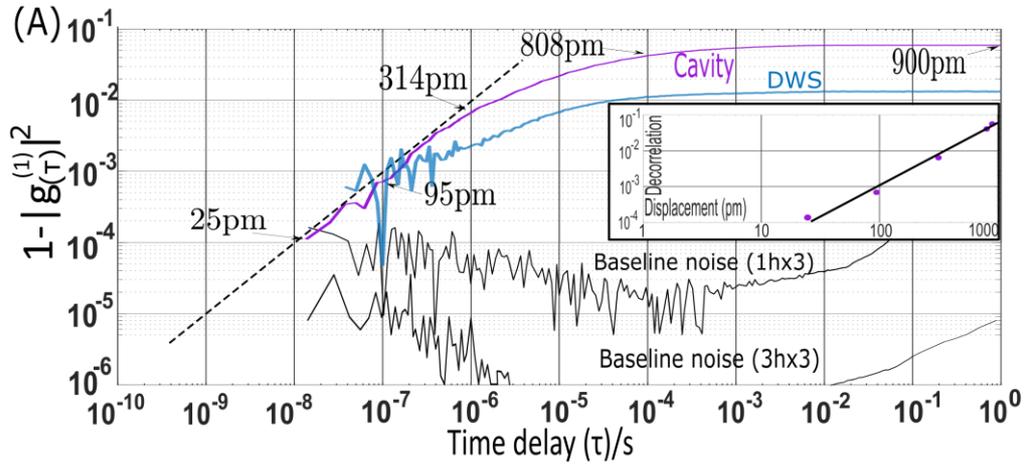

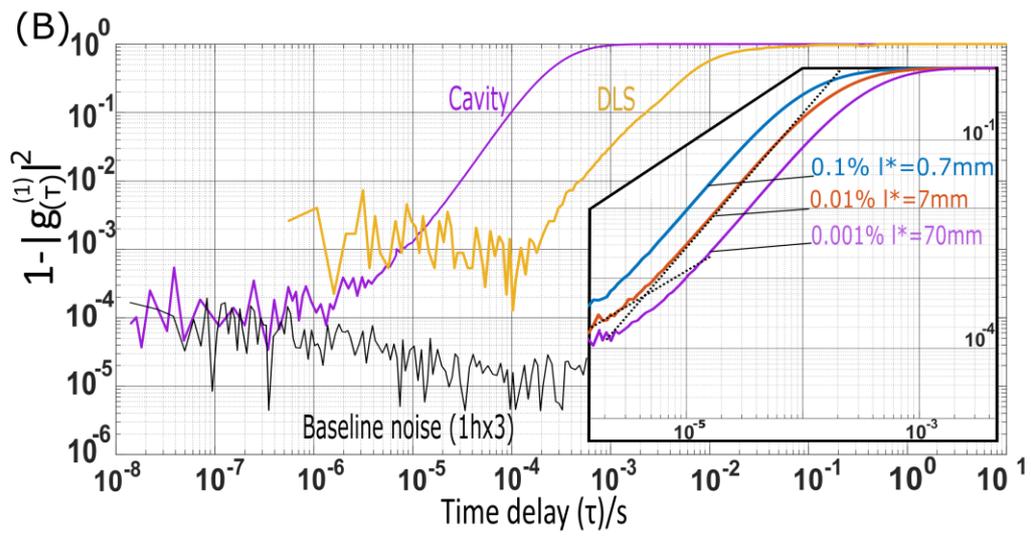

**Fig. 4.** Interferometric response to intracavity light scattering perturbations. Data averaged from N=3 repetitions of 1 h acquisition times, laser power is 300 mW for all experiments. The cavity has a volume $V_c = 118 cm^3$, a wall area $\Sigma_c = 134 cm^2$, and a mean chord length $\bar{\Lambda}_c = 3.53 cm$. The effective gain is $g_e \approx 1700$. (A) Plot (purple line) of the speckle intensity decorrelation $1 - |g^{(1)}(\tau)|^2$ measure inside the cavity and caused by the thermal fluctuations of a jammed monodisperse emulsion of polydimethyl siloxane (PDMS) oil-in-water, with average droplet radius 459±15 nm, volume fraction 0.738, transport mean free path $l^*_{s1} = 0.25 mm$. The sample volume is $V_{s1} = 0.785 cm^3$, the surface area $\Sigma_{s1} = 4.71 cm^2$, and the mean chord length $\bar{\Lambda}_{s1} = 0.67 cm$. The arrows point to specific values of the root-mean square (*rms*) displacement of the droplets computed from (13). Black curves represent the standard error $\sigma_{|g^{(1)}|^2}(\tau)/\sqrt{N}$ measured with the empty unperturbed cavity from three 1 h acquisitions (rescaled by $\sqrt{N}$ compared to Fig.2b). (insert) Measured decorrelation amplitude as a function of the *rms* displacement amplitude in picometers, with the solid line showing the linear response. The (blue line) represents a measurement of the same sample performed under the same experimental conditions on a self-made Diffusing Wave Spectroscopy setup.

(B) Plots of the intensity decorrelation signal $1 - |g^{(1)}(\tau)|^2$ for water suspensions of 20μm diameter Polymethyl methacrylate (PMMA) spheres in water. The response for a $10^{-5}$ volume fraction (purple line) inside the cavity is compared to the baseline noise amplitude (black line) of an empty unperturbed cavity. The (yellow line) represents a measurement of the same sample performed under the same experimental conditions using a commercial Dynamic Light Spectroscopy instrument. (insert) Decorrelation signal obtained with the cavity for three volume fractions, $10^{-3}$ (blue, $l^* = 0.7mm$), $10^{-4}$ (red, $l^* = 7mm$) and $10^{-5}$ (purple, $l^* = 70mm$). All samples have a volume $V_{s2} = 19.6cm^3$, a surface area $\Sigma_{s2} = 55cm^2$, and a mean chord length $\bar{\Lambda}_{s2} = 1.43cm$.